\documentclass[letterpaper, 10 pt, conference]{ieeeconf}  % Comment this line out if you need a4paper

\usepackage{ghasemi}
\usepackage{cite}
\usepackage{mathrsfs}
\usepackage{amsmath}
\usepackage{amsfonts}
\usepackage{amssymb,latexsym}
\usepackage[psamsfonts]{eucal}
\usepackage{graphicx}
\usepackage{epsfig}
\usepackage{psfrag}
\usepackage{subfigure}
\usepackage{indentfirst}
\usepackage{setspace}
\usepackage{epstopdf}
\usepackage{pict2e}
\usepackage{graphicx}
\usepackage{xcolor}
\usepackage{multicol}
\usepackage{mathtools, cuted}
\usepackage{lipsum, color}
\usepackage{soul}

\usepackage{comment}

\IEEEoverridecommandlockouts                              % This command is only needed if 
\title{\LARGE \bf
Impedance Modulation for Negotiating Control Authority in a Haptic Shared Control Paradigm}

\author{Vahid Izadi$^{1}$, Akshay Bhardwaj$^{2}$, and Amir H. Ghasemi$^{1}$% <-this % stops a space
\thanks{$^{1}$ Vahid Izadi and Amir H. Ghasemi are with the Department of Mechanical Engineering, the University of North Carolina at Charlotte, Charlotte, NC, 28223, {\tt\small vizadi@uncc.edu},{\tt\small ah.ghasemi@uncc.edu}}%
\thanks{$^{2}$ Akshay Bhardwaj is with the Department of Mechanical Engineering, University of Michigan, Ann Arbor, MI, 48109, {\tt\small akshaybh@umich.edu }}%%
}

\begin{document}

\maketitle
\thispagestyle{empty}
\pagestyle{empty}

%%%%%%%%%%%%%%%%%%%%%%%%%%%%%%%%%%%%%%%%%%%%%%%%%%%%%%%%%%%%%%%%%%%%%%%%%%%%%%%%
\begin{abstract}
 Communication and cooperation among team members can be enhanced significantly with physical interaction.  Successful collaboration requires the integration of the individual partners' intentions into a shared action plan, which may involve a continuous negotiation of intentions and roles. This paper presents an adaptive haptic shared control framework wherein a human driver and an automation system are physically connected through a motorized steering wheel. By virtue of haptic feedback, the driver and automation system can monitor each other actions and can still intuitively express their control intentions. The objective of this paper is to develop a systematic model for an automation system that can vary its impedance such that the control authority can transit between the two agents intuitively and smoothly. To this end, we defined a cost function that not only ensures the safety of the collaborative task but also takes account of the assistive behavior of the automation system. We employed a predictive controller based on modified least square to modulate the automation system impedance such that the cost function is optimized. The results demonstrate the significance of the proposed approach for negotiating the control authority, specifically when humans and automation are in a non-cooperative mode. Furthermore, the performance of the adaptive haptic shared control is compared with the traditional fixed automation impedance haptic shared control paradigm. 
\end{abstract}

%%%%%%%%%%%%%%%%%%%%%%%%%%%%%%%%%%%%%%%%%%%%%%%%%%%%%%%%%%%%%%%%%%%%%%%%%%%%%%%%
\section{INTRODUCTION}
Haptic shared control is an emerging research topic with a wide range of applications in the areas such as smart manufacturing, autonomous driving, rehabilitation, health-care, education, and training \cite{agah2000human, albu2005physical, beyl2011safe, boehm2016architectures, ghasemi2016role,ghasemi2018adaptive, ghasemi2018game,ghasemi2019shared, vitiello2013neuroexos, bhardwaj2020s, izadi2019determination}. Traditionally, interactive robots were designed to act mainly as reactive followers where the robot (with some level of autonomy) followed the human's commands \cite{ gillespie1998virtual, haanpaa1997advanced, park2001virtual}. However, this type of master-servant arrangement does not capture the sense of partnership \cite{ ikeura2002optimal, ikeura1997variable, arai2000human} that we mean when we speak of two humans cooperatively moving a piece of furniture. A robot as a pro-active partner, also called a co-robot, should be capable of monitoring human actions, as well as communicating its behavior, and even negotiating and exchanging roles with a human partner\cite{reed2008physical, groten2009experimental}.  These criteria give rise to a set of fundamental questions that we aim to answer in this research. For instance, (1) what are the interaction models between a human and co-robot in a Haptic Shared Control framework? (2) Knowing the interaction models, how should a co-robot facilitate exchanging roles dynamically? (3) What strategies should a co-robot take to create consensus with a human while also guaranteeing the safety and performance of the task? Moreover, (4) how may uncertainty in the behavior of the human-operator affect these interaction models and consensus models?   

To seamlessly combine the commands of a human operator and automation system, most of the existing efforts have been devoted to designing an interface so that the human's high-level commands (human's intentions) can be exploited. Specifically, in the context of physical human-robot interaction, force/torque sensors are embedded in the haptic interface for recognizing human intentions and consequently adjusting the robot's behavior. %\cite{Omally}. 
However, when the automation system and human operator simultaneously interact with each other (especially in an uncertain environment), the force sensors measurements are insufficient for determining the human's intents. To resolve this issue, we propose to measure the human's impedance (stiffness of the muscles) as a potential indicator for determining how a human operator dynamically exchanges his role (leader/follower) within a collocative task. We argue if the roles of the two agents are agreed upon, then an appropriately timed nudge from one agent can be interpreted by the other and followed or optionally hindered.

To solve an optimal control problem, several methods exist such as least squares (LS), linear programming (LP), and quadratic programming (QP). The constraints on the control signal or the states play the main role in the solutions of such problems. In this paper, to achieve a non-negative value for the automation's impedance with a computationally inexpensive method, a modified version of least squares is provided.

The outline of this paper is as follows. Section II presents the basics of adaptive haptic shared control paradigm. Section III presents the basics of a controller that adaptively modulate the automation's impedance. 
Section IV presents numerical results followed by Section V which presents the conclusions and future work.

\section{Adaptive Haptic Shared Control Framework}

\begin{figure*}[htbp]
    \centering
    \includegraphics[width=\textwidth]{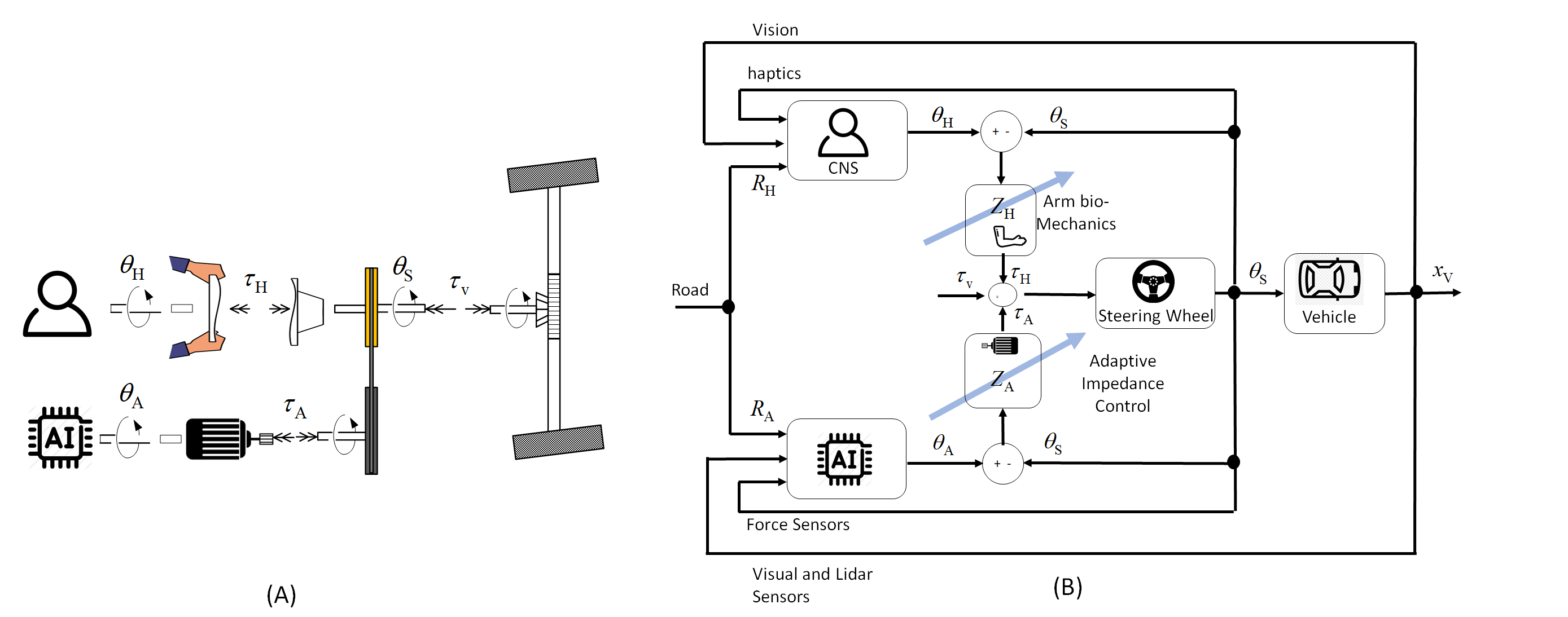}
    \caption{(A) A general model of control sharing between driver and automation. (B) A block diagram is laid out to highlight the
interaction ports between subsystems.}
    \label{HRI}
\end{figure*}

Figure \ref{HRI} shows a  schematic of an adaptive haptic shared control.  Three entities each impose a torque on the steering wheel: a human driver through his hands, an automation system through a motor, and the road through the steering linkage.  The driver model consists of a  cognitive controller, coupled with a biomechanics subsystem. In Figure \ref{HRI}, $\theta_{\rm H}$ represents the driver’s intent, which is an output of the cognitive controller. Impedance $Z_{\rm H}$ represents the biomechanics of the driver's arm, which is backdriveable (the driver impedance is not infinite). To indicate that driver impedance varies with changes in grip on the steering wheel, use of one hand or two, muscle co-contraction, or posture changes, we have drawn an arrow through $Z_{\rm H}$. In this paper, the driver and automation system are both shown with a similar structure. Specifically, the automation system is modeled as a higher-level controller (AI) coupled with a lower-level impedance controller. The focus of this paper is on the design of a backdrivable impedance $Z_{\rm A}$. That is, the automation is not designed to behave as an ideal torque source. Rather,  the automation imposes its command $\theta_{\rm A}$ through an impedance $Z_{\rm A}$, which can be varied under control of the automation to express the automation's current level of control authority. Furthermore, the reference signals $R_{\rm H}$ and $R_{\rm A}$ represent the goals of the driver and the automation system, respectively. It should be noted that these goals may not necessarily be the same, which is when the negotiation of control authority becomes important. From the model presented in Figure \ref{HRI}, it follows that the position of the steering wheel $\theta_{\rm S}$ is a function of the human’s intent $\theta_{\rm H}$, automation’s intent $\theta_{\rm A}$, and the road feedback torque $\tau_{\rm V}$ that results from the forces arising on the steering rack due to tire-road interaction \cite{bhardwaj2019estimating}. For a certain impedance, the steering wheel angle is 
\begin{align}
{\theta_{SW}}(s)=\frac{Z_{\rm H}(s) \theta_{\rm H}(s)+ Z_{\rm A}(s) \theta_{\rm A}(s)+ \tau_{\rm V}(s) }{ J_{\rm eq}+B_{\rm SW}s+ Z_{\rm H}(s)+ Z_{\rm A}(s)}\label{thetaS}
\end{align}
where $J_{\rm eq}=J_{\rm SW}+J_{\rm H}+J_{\rm A}$, where $J_{\rm SW}$ is the steering wheel inertia, $J_{\rm H}$ is the inertia of human's hand and $J_{\rm A}$ is the inertia of automation motor \cite{ghasemi2019shared, boehm2016architectures} .

It follows from (\ref{thetaS}) by changing his/her impedance (increasing $Z_{\rm H}$ by co-contracting muscles), the
human driver can increase his control authority. Likewise, the automation can be designed such that by imposing higher impedance, and it requests a higher control authority. In this paper, we define the level of authority as a relationship between the human's impedance and automation's impedance.

To present how impedance may evolve in time, we introduce  the following dynamic models:
\begin{align}
    \dot Z_{\rm H}(t)&=\alpha_{\rm H} Z_{\rm H}(t)+\beta_{\rm H}\Gamma_{\rm H}(t) \label{ZH}\\
    \dot Z_{\rm A}(t)& =\alpha_{\rm A} Z_{\rm A}(t)+\beta_{\rm A}\Gamma_{\rm A}(t) \label{ZA}
\end{align}
where $Z_{\rm H}=[B_{\rm H} \ K_{\rm H} ]^{\rm T}$  and $K_{\rm H}$ and $B_{\rm H}$ are the stiffness and damping associated with humans' biomechanics; $Z_{\rm A}=[B_{\rm A} \ K_{\rm A} ]^{\rm T}$ and $K_{\rm A}$ and $B_{\rm A}$ are the stiffness and damping associated with the motor's lower-level proportional-derivative controller;  $\Gamma_{\rm H}=[ \Gamma_{\rm bH}(t)\ \Gamma_{\rm kH}(t)]^{\rm T}$ is the human’s control action for modulating his impedance and $\Gamma_{\rm A}=[ \Gamma_{\rm bA}(t)\ \Gamma_{\rm kA}(t)]^{\rm T}$ is the automation’s control input for modulating its impedance. Additionally, 
\begin{align}\label{al_bet}
\alpha_{\rm H}=&\left[ \begin{matrix}
            \alpha_{\rm bH} & 0 \\ 0 & \alpha_{\rm kH}\end{matrix}\right], \quad 
            \beta_{\rm H}= \left[ \begin{matrix}
            \beta_{\rm bH} & 0 \\ 0 & \beta_{\rm kH}\end{matrix}\right] \nn 
            \\
          \alpha_{\rm A}=&  \left[ \begin{matrix}
            \alpha_{\rm bA} & 0 \\ 0 & \alpha_{\rm kA}\end{matrix}\right],
   \quad \beta_{\rm A}= \left[ \begin{matrix}
            \beta_{\rm bA} & 0 \\ 0 & \beta_{\rm kA}\end{matrix}\right]
\end{align}
where $\{\alpha_{\rm bH}, \alpha_{\rm kH}, \alpha_{\rm bA}, \alpha_{\rm kA}, \beta_{\rm bH}, \beta_{\rm kH}, \beta_{\rm bA}, \beta_{\rm kA}\}$ are constant parameters.   This formulation captures how impedance evolves in time. Ideally, to determine an optimal behavior for the automation system, optimization should be performed over all control signals  of the automation system (i.e., $\theta_{\rm A}, \Gamma_{\rm A}$). However, the focus of this paper is  to determine $\Gamma_{\rm A}$  as means for allocating the level of authority between the driver and the automation system.

\section{Impedance Modulation Controller Design}

In this section, we present a predictive controller for modulating the automation impedance such that the assistive behavior of the automation system improves while the safety of the task is also guaranteed.  

The discrete-time model of the impedance dynamics (\ref{ZH}) and (\ref{ZA}) using the Forward Euler method would be
    %\mathbf  Z_{\rm H}(k+1)=\mathbf \alpha_{\rm H} \mathbf Z_{\rm H}(k)+\mathbf \beta_{\rm H}\mathbf \Gamma_{\rm H}(k)  \\
      %  \mathbf  Z_{\rm A}(k)=\mathbf \alpha_{\rm A} \mathbf Z_{\rm A}(k)+\mathbf \beta_{\rm A}\mathbf \Gamma_{\rm A}(k)
      \begin{comment}
 \begin{align}
    \left[\begin{matrix}
     Z_{\rm H}(k+1)\\ 
    %K_{\rm H}(k+1)
    \end{matrix}\right]=\left[ \begin{matrix}
            \widetilde{\alpha}_{\rm H} B_{\rm H}(k)+\widetilde{\beta}_{\rm H}\Gamma_{\rm BH}(k)\\ 
            \widetilde{\alpha}_{\rm H} K_{\rm H}(k)+\widetilde{\beta}_{\rm H}\Gamma_{\rm KH}(k)
    \end{matrix}\right]\label{ZH_discreet_mat}\\ \left[\begin{matrix}
    B_{\rm A}(k+1)\\ 
    K_{\rm A}(k+1)
    \end{matrix}\right]=\left[ \begin{matrix}
            \widetilde{\alpha}_{\rm A} B_{\rm A}(k)+\widetilde{\beta}_{\rm A}\Gamma_{\rm BA}(k)\\ 
            \widetilde{\alpha}_{\rm A} K_{\rm A}(k)+\widetilde{\beta}_{\rm A}\Gamma_{\rm KA}(k)
    \end{matrix}\right]\label{ZA_discreet_mat}
\end{align}
\end{comment}
\begin{align}
    Z_{\rm H}(k+1)&=\widetilde \alpha_{\rm H} Z_{\rm H}(k)+\widetilde \beta_{\rm H}\Gamma_{\rm H}(k+1) \label{ZHD}\\
     Z_{\rm A}(k+1)& =\widetilde \alpha_{\rm A} Z_{\rm A}(k)+\widetilde \beta_{\rm A}\Gamma_{\rm A}(k+1) \label{ZAD}
\end{align}
where $ \widetilde{\alpha}_{\rm H} = ({I-T_{\rm s} {\alpha}_{\rm H}})^{-1}$,    $\widetilde{\beta}_{\rm H} = \widetilde{\alpha}_{\rm H}{T_{\rm s} {\beta}_{\rm H}}$, $ \widetilde{\alpha}_{\rm A} = ({I-T_{\rm s} {\alpha}_{\rm A}})^{-1}$,    $\widetilde{\beta}_{\rm A} = \widetilde{\alpha}_{\rm A}{T_{\rm s} {\beta}_{\rm A}}$, and $T_{\rm s}$ is the sampling time.

Now let us define the vector $\Theta_i(k) = [\dot{\theta_i}{(k)} \ \ {\theta_i}{(k)}]$ where $i\in \{\rm SW, \rm H,\rm A\}$. Note that
\begin{align}
    \dot{\theta_i}{(k)}=  \frac{{\theta_i}{(k)}-{\theta_i}{(k-1)}}{T_{\rm s}} \label{dteta_discreet}
\end{align} 
Next, let us define a cost function $J(k)$ in the form of 
\begin{align}\label{Cost_func_org_modified2}
\vspace{-1 cm}
    \min_{\Gamma_{\rm A}}  J(k) =  \sum_{j=k+1}^{k+N_p}\{\| |Z_{\rm H}(j)^{T}\Theta_{\rm H}(j)^{T} + Z_{\rm A}(j)^{T} \Theta_{\rm A} (j)^{T}|\\ -\varepsilon (j) \|\nn +\|Z_{\rm H}(j)^{T} \Theta_{\rm H}(j)^{T}-Z_{\rm A}(j)^{T} \Theta_{\rm A}(j)^{T}\|\}
\end{align}
The first term of the cost function is to ensure safe steering. Specifically, we define $\varepsilon$ as a minimum required torque that can guarantee the safe maneuver. In this paper, we assume $\varepsilon$ is known. The second term of the cost function is to minimize the disagreement between a human driver and the automation system. The steering angle, $\theta_{\rm S}$ and its rate of change $\dot{\theta}_{\rm S}$ can be directly measured from the sensor. Further, we assume that $Z_{\rm H}$ and $\Theta_{\rm H}$ are known and can be measured. 

The goal in the cost function is to determine $\Gamma_{\rm A}$ such that the cost function $J$ is minimized. To this end, $Z_{\rm A}^{T}{\Theta_{\rm A}^{T}}$ can be presented as
\begin{align}
\begin{aligned}
    Z_{\rm A}{(k)}^{T}{\Theta_{\rm A}}{(k)}^{T} =
    {B_A}(k)\Big [\frac{{\theta_{\rm A}}{(k)}-{\theta_{\rm A}}{(k-1)}}{T_{\rm s}}\Big]   \\+{K_{\rm A}}(k){\theta_{\rm A}}{(k)} \label{tauA}
    \end{aligned}
\end{align}
%By replacing $B_{\rm A}$ and $K_{\rm A}$ from Eq. \ref{tauA}, we will have:
Eq. \ref{tauA} can be rewritten as 
\begin{align}\label{tauAmatrix2}
   Z_{\rm A}{(k)}^{T}{\Theta_{\rm A}}{(k)}^{T} = (\Phi(k) +\Psi (k))
   \left[\begin{matrix}
            \theta_{\rm A}(k)\\ 
            \theta_{\rm A}(k-1)
    \end{matrix}\right],
\end{align}
where
\begin{align}
    {\Phi}(k) = \widetilde{\alpha}_{\rm A}\begin{bmatrix}
                \frac{{B_{\rm A}}(k-1)}{T_{\rm s}}+{K_{\rm A}}(k-1) & 
               -\frac{{B_A}(k-1)}{T_{\rm s}}
    \end{bmatrix}\label{PhiAmatrix}
\end{align}
\begin{align}
    \Psi(k) = \widetilde{\beta}_{\rm A} \left[\begin{matrix}
               \frac{{\Gamma_{\rm BA}}(k)}{T_{\rm s}}+{\Gamma_{{\rm KA}}}(k) & 
               -\frac{{\Gamma_{\rm BA}(k)}}{T_{\rm s}}
    \end{matrix}\right]\label{Psi_A_matrix}
\end{align}

The $\Phi$ and $\Psi$ represent modified mechanical impedance and control action vectors, respectively. By propagating the automation torque for the next time steps until $N_p$ step, the $\Phi$ and $\Psi$ vectors will move forward in the time. 
% \begin{comment}
% \begin{align}
%   Z_{\rm A}{(k+1)}&{\theta_{\rm A}}{(k+1)} =\nonumber\\
%   \left \{  \widetilde{\alpha}_A[\Phi{(k)} +\Psi{(k)}] +\Psi{(k+1)}\right \}
%   &\begin{bmatrix}
%             \theta_{\rm A}(k+1)\\ 
%             \theta_{\rm A}(k)\\ 
%             \theta_{\rm A}(k-1)
%     \end{bmatrix}\label{tau_A_matrix_k1}
% \end{align}
% \begin{align}
%   Z_{\rm A}{(k+n)}{\theta_{\rm A}}{(k+n)} =\nonumber\\\nonumber
%   \left \{(\widetilde{\alpha}_A)^n[\Phi{(k)} +\Psi{(k)}] +(\widetilde{\alpha}_A)^{n-1}\Psi{(k+1)}+...+\Psi{(k+n)}\right \}\\ \nonumber
%   \begin{bmatrix}
%             \theta_{\rm A}(k+n)\\ 
%             \theta_{\rm A}(k+n-1)\\ 
%             \theta_{\rm A}(k+n-2)
%     \end{bmatrix}\label{tau_A_matrix_kn}\\
% \end{align}
% \end{comment}
Let us create the prediction matrices for $N_p$ steps. From Equation (\ref{tauAmatrix2}), we obtain the following $Z_{\rm A}{(k+1)}^{T}{\Theta_{\rm A}}{(k+1)}^{T}$ for step $(k+1)$

\begin{align}\label{tauAmatrix3}
\begin{aligned}
   Z_{\rm A}{(k+1)}^{T}{\Theta_{\rm A}}{(k+1)}^{T} = (\widetilde{\alpha}_A \lbrack \Phi(k) +\Psi (k) \rbrack \\ + \Psi (k+1))
   \left[\begin{matrix}
            \theta_{\rm A}(k)\\ 
            \theta_{\rm A}(k-1)
    \end{matrix}\right]
    \end{aligned}
\end{align}
Propagating further to $(k+N_p)$ index we obtain

\begin{align}\label{tauAmatrix4}
\begin{aligned}
   Z_{\rm A}{(k+N_p)}^{T}{\Theta_{\rm A}}{(k+N_p)}^{T} = ((\widetilde{\alpha}_A)^{N_p} \lbrack \Phi(k) +\Psi (k) \rbrack \\+ (\widetilde{\alpha}_A)^{N_p-1} \Psi (k+1) + \cdots + \Psi (k+N_p))
   \left[\begin{matrix} \theta_{\rm A}(k)\\\theta_{\rm A}(k-1) \end{matrix}\right]
    \end{aligned}
\end{align}
The prediction matrix $\mathbf{Z}_{\rm A}^{T} \boldsymbol\theta_{\rm A}^{T}$ can then be written as follows

\begin{align}
    \mathbf{Z}_{\rm A}^{T} \boldsymbol\theta_{\rm A}^{T}=\overline{\mathbf{\Theta}} \mathbf{\Omega (\mathbf{{\Phi, \Psi} })},
\end{align}
where 
\begin{align}
&    \mathbf{Z}_{\rm A}^{T} \boldsymbol\Theta_{\rm A}^{T}= \begin{bmatrix}
            Z_{\rm A}{(k+N_p)}^{T}{\Theta_{\rm A}}{(k+N_p)}^{T}\\ 
            \vdots\\ 
            Z_{\rm A}{(k+1)}^{T}{\Theta_{\rm A}}{(k+1)}^{T}\\
            Z_{\rm A}{(k)}^{T}{\Theta_{\rm A}}{(k)}^{T}
    \end{bmatrix}\label{A_LS_func}\\
& \overline{\mathbf{\Theta}}^{\rm T}=\begin{bmatrix}
 {\Theta_{\rm A}}{(k+N_p)}&  \cdots &  0& 0\\ 
 {\Theta_{\rm A}}{(k+N_p-1)}&  \cdots &  0& 0\\ 
 \vdots &  \ddots &  \vdots & \vdots\\ 
 0&  \cdots & {\Theta_{\rm A}}{(k+1)} & 0\\ 
 0&  \cdots & {\Theta_{\rm A}}{(k)} & 0\\ 
 0&  \cdots &  0& {\Theta_{\rm A}}{(k)}\\       
 0&  \cdots &  0& {\Theta_{\rm A}}{(k-1)}
\end{bmatrix}\end{align}
\begin{align}
    &\mathbf{\Omega (\mathbf{{\Phi, \Psi} })}\nn =\\&\begin{bmatrix}
            \left \{\widetilde{\alpha}_A^{N_{\rm p}}[\Delta(k)] +\widetilde{\alpha}_A^{{N_{\rm p}}-1}\Psi{(k+1)}+...+\Psi{(k+{N_{\rm p}})}\right \}^{\rm T}\\ 
            \vdots\\ 
            \left \{  \widetilde{\alpha}_A[\Delta(k)] +\Psi{(k+1)}\right \}^{\rm T}\\
            \left \{ \Delta(k) \right \}^{\rm T}
    \end{bmatrix}
\end{align}
where $\Delta(k)=\Phi{(k)} +\Psi{(k)}$.
%\end{strip}

According to the second term of (\ref{Cost_func_org_modified2}), in the ideal model, the value of $Z_{\rm H}^T\Theta_{\rm H}^T$ will be equal to $Z_{\rm A}^T\Theta_{\rm A}^T$. On the other hand, in (\ref{A_LS_func}), the amount of automation control action at a time step $k$ can be determined by using methods like linear programming (LP), quadratic programming (QP) and least square (LS). As it can be seen in the (\ref{A_LS_func}), the solution from the optimal solver will give us the summation of $\Phi$ and $\Psi$. In the LP, QP, and LS methods, it is possible to have a negative value. This means, the mechanical link on the automation side is disconnected (has zero impedance). In this paper, a modified version of the LS method is used to solve the cost function with a non-negative solution (non-negative impedance).The modified least-square is an LS optimization problem which is subjected to non-negativity constraints. The procedure to implement these constraints is to solve the corresponding unconstrained LS problem and then overwrite any negative values with zeros. %Furthermore, we used a modified version of NNLS to solve the cost function and reduce the computational burden. In the following section, the MNNLS method is presented.

\section{Results}
In this section, we present a series of simulation results to show the effectiveness of the proposed adaptive haptic shared control (HSC) for improving the collaboration between the human driver and the automation system. We specifically consider two modes of interactions: cooperative and non-cooperative. A cooperative mode is when $R_{\rm H}$ and $R_{\rm A}$ (see Fig. \ref{HRI}) have the same sign and non-cooperative mode is when the sign of $R_{\rm H}$ and $R_{\rm A}$ is different. In this paper, we assume $\theta_{\rm H}$ and $\theta_{\rm A}$, which are the outputs of a higher-level controller (e.g., MPC-based controller) are known and given. We aim to determine an optimal $\Gamma_{\rm A}$ such that the cost function $J$ defined in (\ref{Cost_func_org_modified2}) is minimized. 
%the results from the numerical simulations are demonstrated. The human and robot impedance parameters ($B$ and $K$ values) are demonstrated for different cooperative and non-cooperative modes. The reference value for path tracking is defined by a higher-level controller in the brain on an AI processor. Then, these reference values at each time define the cooperativeness in the human-robot framework.
For all the simulation results presented below,  the steering wheel inertia is $J_{\rm SW}=0.1$ N.m/rad/$s^2$ and damping coefficient for the steering wheel is  $B_{\rm SW}=0.01$ N.m/rad/s. Also, the human's hand inertia is $J_{\rm H}=0.001$ N.m/rad/$s^2$  \cite{yu2014human} and the motor inertia is $J_{\rm A}=J_{\rm H}$. The sampling time is $T_{\rm s}= 0.1$ second and the control and estimation horizon is $N_{\rm p}=20$.  The matrices $\alpha_{\rm H}$, $\alpha_{\rm A}$, $\beta_{\rm H}$, and $\beta_{\rm A}$ defined in Equation (\ref{al_bet}) are all assumed to be identity matrices. The road torque feedback $\tau_{\rm V}$ is assumed to be zero. 

% According to the proposed approach in this paper, in the cooperative mode, the automation system aims to imitate the impedance of the human while it is equal to a non-zero value. In the case of zero impedance from the driver side, the automation system based on the first term of the cost function generates non-zero impedance. }

\begin{figure}[htbp]
\vspace{.1 cm}
\centering
\includegraphics[width=0.48\textwidth]{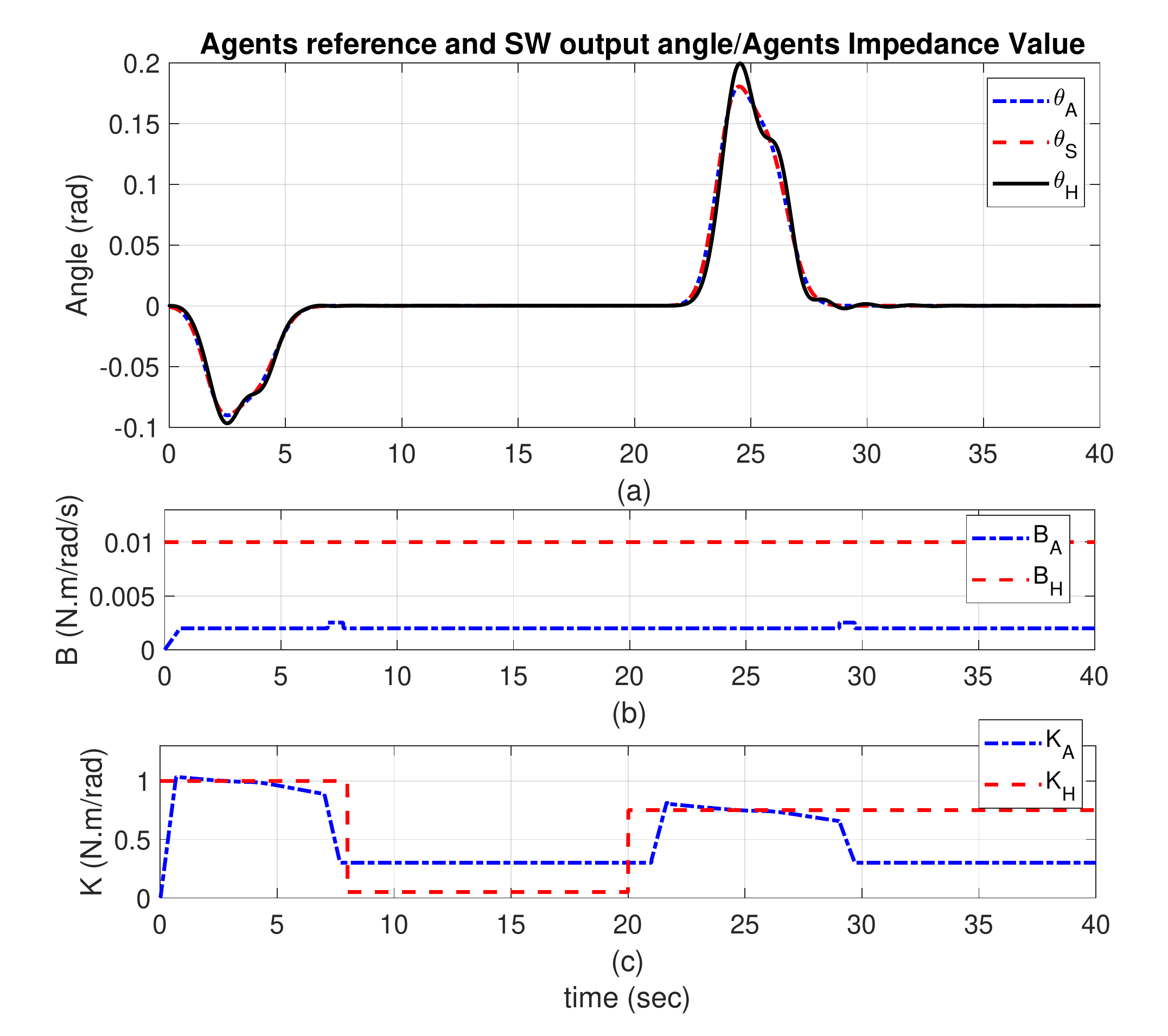}
\caption{Cooperative mode of integration between a human driver and an automation system with adaptive HSC.}
\label{fig3}
\vspace{-.4 cm}
\end{figure}

Figure \ref{fig3} demonstrates a cooperative mode of interaction between a human driver and an automation system when both driver and automation intentions have the same sign (${\rm sgn}(\theta_{\rm H})={\rm sgn}(\theta_{\rm A})$). The human's impedance parameters are shown with red dashed lines. We consider a scenario where the human's stiffness is adaptively changing through time. Specifically, the initial value of stiffness $K_{\rm H}$ is $1$ N.m/rad and at $t=8$ and $t=20$ seconds and it changes from $1$ N.m/rad to $0.05$ N.m/rad and from $0.05$ N.m/rad to $0.75$ N.m/rad, respectively. The damping is held fixed at $B_H = 0.01$ N.m/rad/s. The Figure \ref{fig3}-c shows that automation tries to match its impedance with the human driver in the cooperative task. 

%In the Figure \ref{fig3}, both driver and automation have same intention and the human agent at 8 and 20 seconds changes his/her impedance from $1 N.red$ to $0.05N.red$ and from $0.05 N.red$ to $0.75N.red$ respectively. In part (c) of Figure \ref{fig3}, the modulated impedance of the automation system is shown, and it imitates the driver behavior in the cooperative task. 

\begin{figure}[htbp]
\vspace{.1 cm}
\centering
\includegraphics[width=0.48\textwidth]{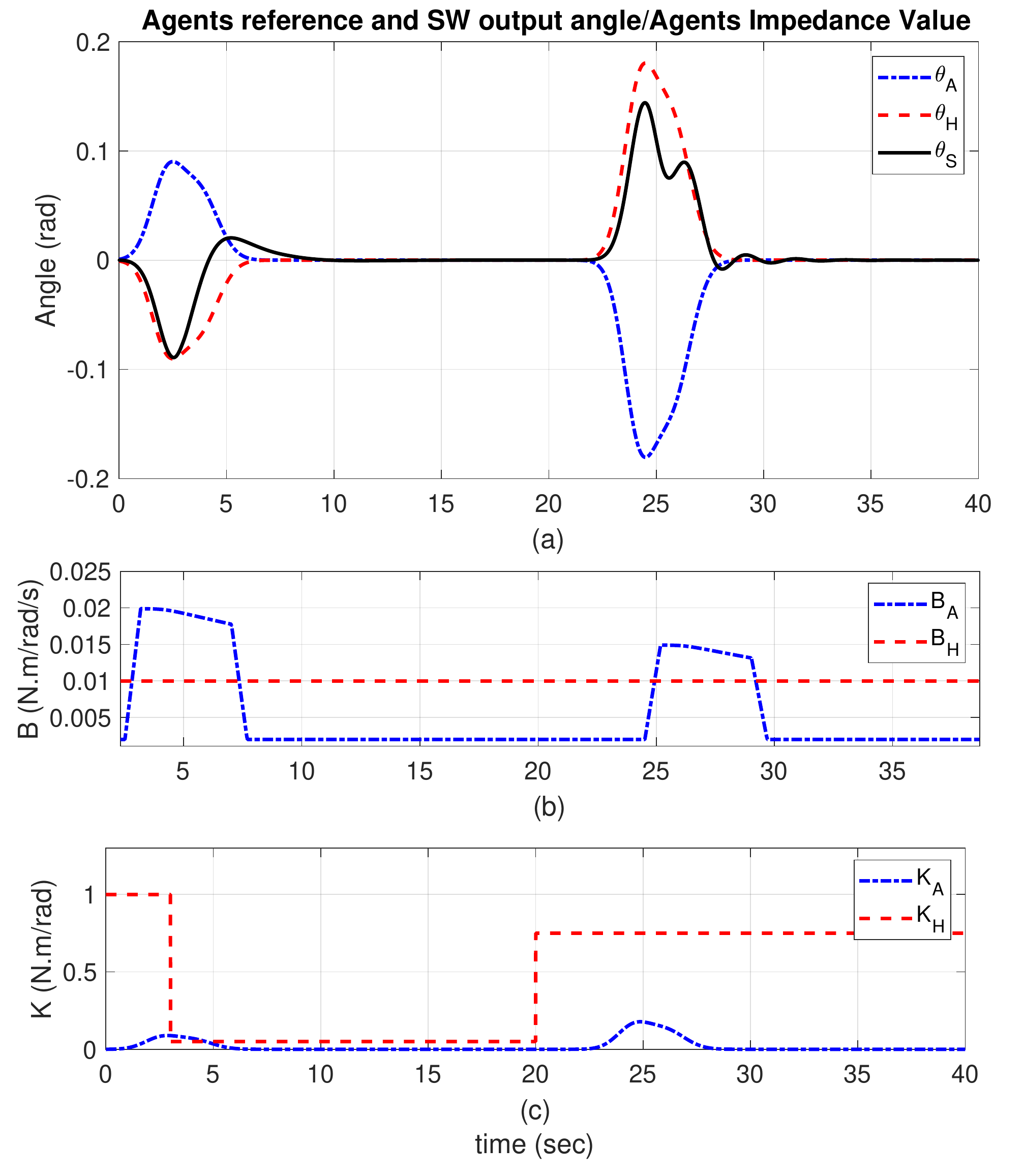}
\caption{Non-cooperative mode of integration between a human driver and an automation system with adaptive HSC.}
\label{fig4}
\vspace{-.4 cm}
\end{figure}

Figure \ref{fig4} demonstrate a non-cooperative mode of interaction between a human driver and an automation system when the driver and automation intentions have different signs (${\rm sgn}(\theta_{\rm H})\neq{\rm sgn}(\theta_{\rm A})$). The initial value of stiffness $K_{\rm H}$ is $1$ N.m/rad and at $t=3$ and $t=20$ seconds and it changes from $1$ N.m/rad to $0.05$ N.m/rad and from $0.05$ N.m/rad to $0.75$ N.m/rad, respectively. The damping is held fixed at $B_H = 0.01$ N.m/rad/s. Similar to Figure \ref{fig3}, the human's impedance dynamically changes with time. However, to reduce the disagreement between the two agents,  the automation system adopts a smaller stiffness. While automation damping $B_A$ also remains low it still changes at the instances of conflict. We are currently investigating the reason behind the variation in damping. %It should be noted that during $8<t<20$ when the human's impedance is low and not sufficient for safely maneuvering the steering wheel (i.e., $|\tau_{\rm H}<\varepsilon|$), then the automation system adopt a higher impedance to ensure the safe maneuver.  

%In a non-cooperative task, the automation system, and the driver have different intentions. Then, in this task, the haptic link of the automation will rest, and the driver is dominant in the mechanical interface. The behavior of the automation system, according to the proposed approach in a non-cooperative task, is shown in Figure \ref{fig4}. As it is obvious, $K_{\rm A} \neq 0$ because of the non-zero term of the cost function ($\varepsilon$). 

\begin{figure}[htbp]
\vspace{0 cm}
\centering
\includegraphics[width=0.50\textwidth]{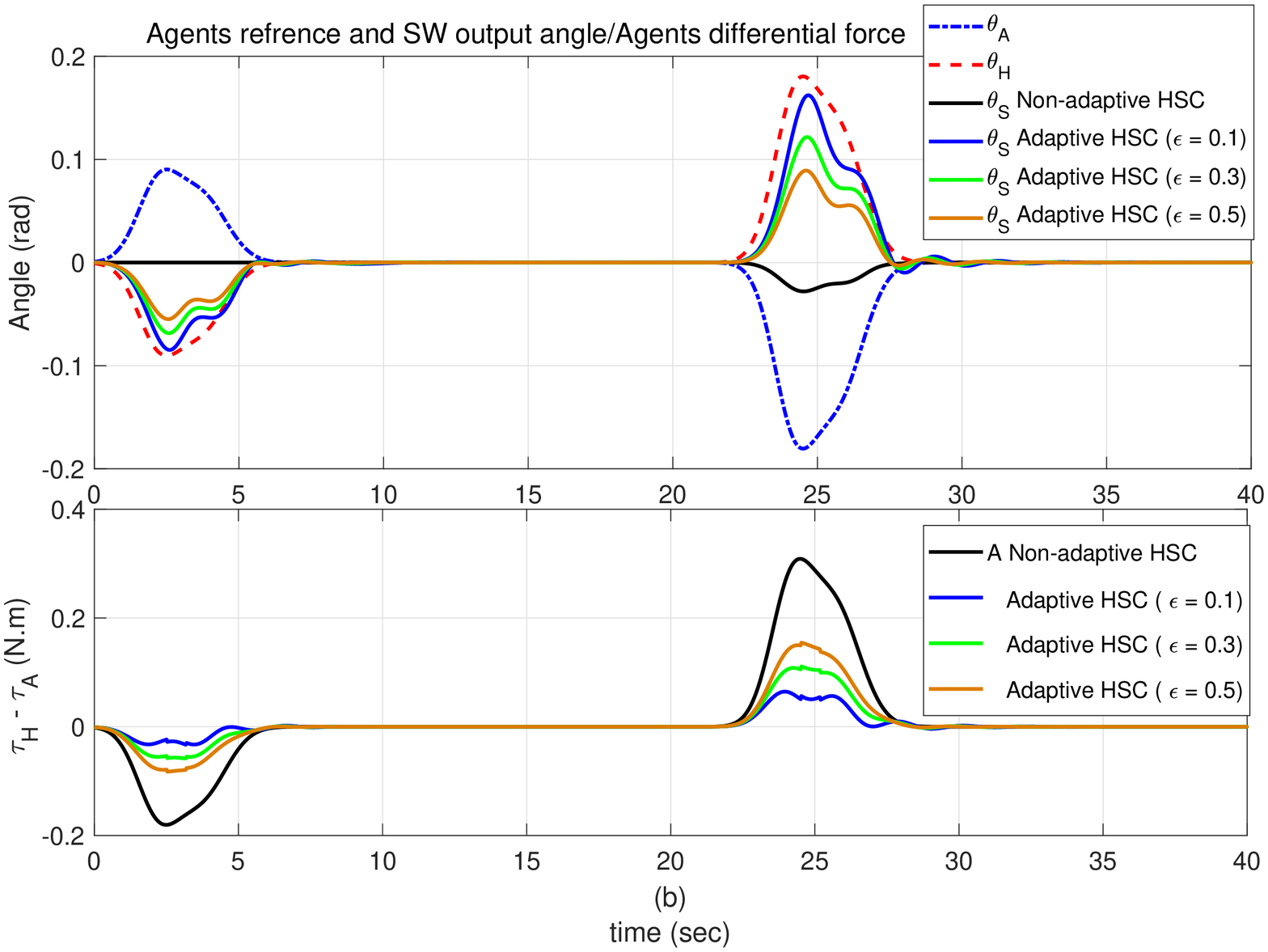}
\caption{The effect of different values of $\varepsilon$ (a) on the steering wheel angle and (b) on the differential torque (conflict) between human and automation.}
\label{fig5}
\vspace{0 cm}
\end{figure}

Figure \ref{fig5} shows how different values of $\varepsilon$ may affect the steering angle as well as the differential torque (fight between the two agents) in the non-cooperative mode. By selecting higher values for $\varepsilon$ the minimum torque required for the safe maneuver is required which results in a bigger differential torque. To ensure the minimum required input for the safe maneuver, there will be a minimum fight between the human and automation system. The amount of fight increases as the amount of minimum required torque increases. %it means that the fight will  and human's and automation's input will cancel each other (lower values of $\theta_{\rm S}$). 

%By setting different values for the $\varepsilon$, the steering wheel angle and the differential torque between driver and automation will be changed. The effect of the different values of $\varepsilon$ in a non-cooperative task on the steering angle and the differential torque is demonstrated in Figure \ref{fig5}. 

\begin{figure}[htbp]
\vspace{0 cm}
\centering
\includegraphics[width=0.52\textwidth]{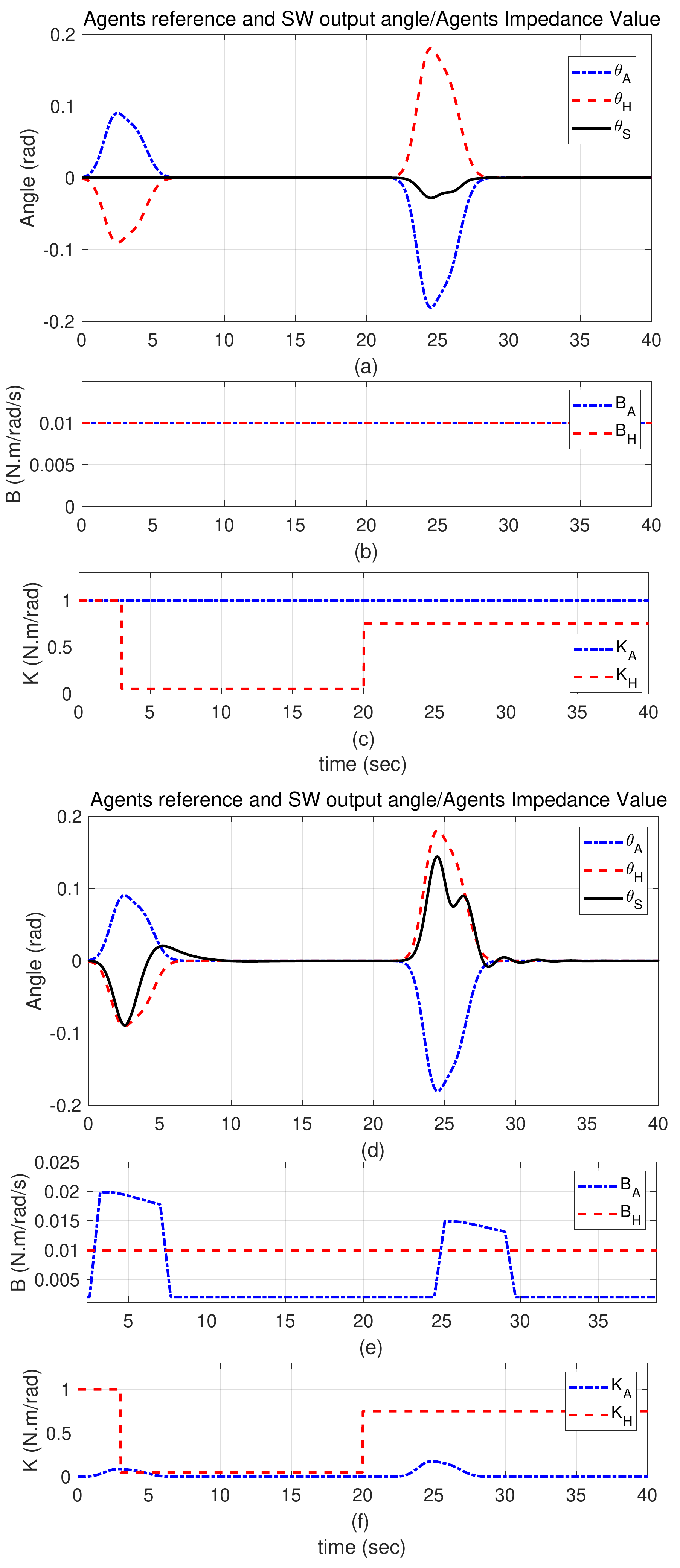}
\caption{Comparison between adaptive and non-adaptive haptic shared control paradigms. Figures (a) to (c) show the reference angle and impedance parameters for the non-adaptive (fixed impedance) haptic shared control paradigm and (d) to (f) show the reference angle and impedance parameters for the adaptive haptic shared control paradigm.}
\label{fig6}
\end{figure}

Next, we compare the performance of a non-adaptive (when the automation impedance does not change as the human's impedance changes) haptic shared control paradigm with an adaptive haptic shared control paradigm in the case of a non-cooperative mode of interaction between the driver and automation. Considering  $\varepsilon=.1$,  (d), (e) and (f) plots of Fig \ref{fig6}  present the impedance and steering angle with impedance modulation (adaptive), while (a), (b) and (c) shows them without impedance modulation (non-adaptive). It follows from Fig \ref{fig6} that in non-adaptive mode, the driver's and automation's control command cancels out, and the steering wheel is almost zero $(\theta_{\rm S}\approx 0)$. However, this issue seems to be solved in adaptive haptic shared control. Additionally, the disagreement between the two agents in the adaptive haptic shared control paradigm is much lower than the non-adaptive haptic mode.

%\color{black}

\section{Conclusions and Future Work}
\begin{figure}[htbp]
\vspace{0 cm}
\centering
\includegraphics[width=0.48\textwidth]{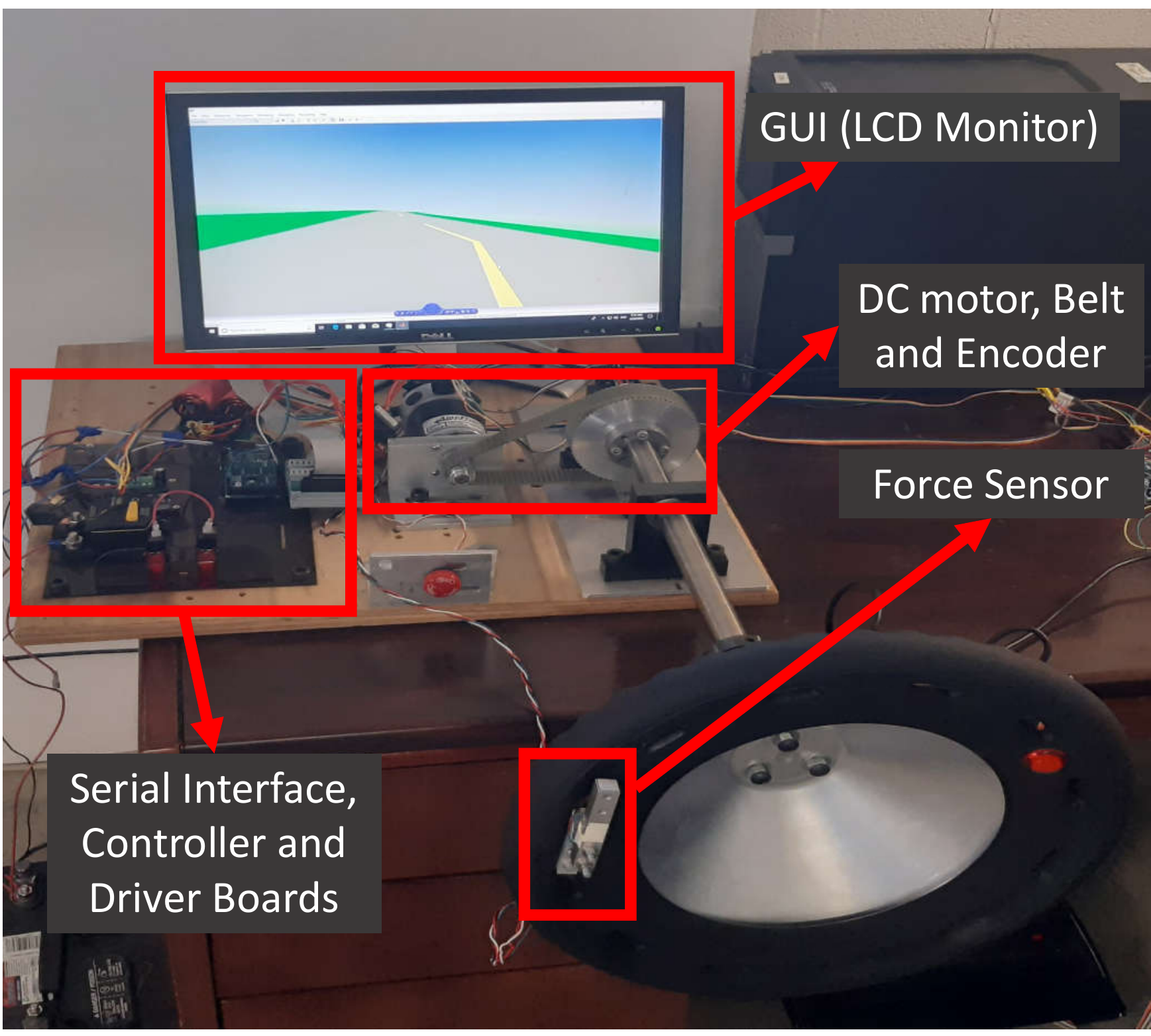}
\caption{Fixed base driving simulator: experimental setup.}
\label{Fig:Setup}
\vspace{0 cm}
\end{figure}

This paper presents the principles of the adaptive haptic shared control paradigm. Specifically, we introduce the impedance modulation as one possible mechanism for negotiation of the control authority. We propose a predictive control approach where the impedance of an automation system is modulated so that the fight between the two agents is minimized while the safety of the system is guaranteed.  
In the future, we plan to extend the outcomes of this research to experimental studies. To this end, we have developed a low-fidelity fixed-base driving simulator (see Fig. \ref{Fig:Setup}). The simulator features a steering wheel that is motorized and a screen that displays a virtual driving environment. %A DC motor (AmpFlow A28-150, Belmont, CA) was coupled to the steering wheel (Speedway 38 cm solid aluminum wheel, Lincoln, NE) through a timing belt with a 72:15 mechanical advantage, making up to 66~Nm torque available to be imposed on the human driver. A 10,000 count per revolution optical encoder (US Digital HB6M, Vancouver, WA) was attached to the steering shaft, and the motor was equipped with a 2048 count per revolution optical encoder (US Digital HB6M). 
The steering wheel is further equipped with a force sensor that is within easy reach of driver's hands. %The virtual driving environment was displayed on a 50~cm LCD Widescreen monitor positioned about 140~cm from the participant. 

So far, we have assumed that driver intent $\theta_{\rm H}$ and driver impedance $Z_{\rm H}$ are available. However, in an actual driving experiment, we would need to estimate these parameters online. In the past, it has been shown that the grip force can be used as a proxy to estimate the driver impedance\cite{yu2014identification, brown2015exploration}. Accordingly, in our driving experiments in the future, we plan to use the grip force sensor measurements to estimate the human's impedance on the fly. Additionally, to have an approximate estimate of the human's intent $\theta_{\rm H}$, we plan to ask the participants to follow a pre-designed path. By integrating this online information about the human's intention (known from a pre-designed path), human's estimated impedance (known from the force sensor), and the automation's intentions (AI's output), we intend to test the performance of the proposed adaptive haptic shared control paradigm on the actual hardware. 

\bibliographystyle{unsrt}     % unsrt plain abbrv
\bibliography{RefACC2019}

\end{document}